\documentclass[%
reprint,
superscriptaddress,
pra,
]{revtex4-2}

\usepackage{graphicx} 
\usepackage{amsmath} 
\usepackage{xcolor}
\usepackage{braket}
\usepackage[colorlinks=true, linkcolor=black, anchorcolor=black, citecolor=blue,
urlcolor = blue]{hyperref}
\usepackage[autostyle]{csquotes}
\newcommand{\mycomment}[1]{}
\usepackage{soul}
\usepackage{lipsum}
\usepackage{braket}

\begin{document}

\title{An Imaging Radar Using a Rydberg Atom Receiver}

\author{William J. Watterson}
\affiliation{ National Institute of Standards and Technology, Boulder, CO 80305, USA}
\author{Nikunjkumar Prajapati}
\affiliation{ National Institute of Standards and Technology, Boulder, CO 80305, USA}
\author{Rodrigo Castillo-Garza}
\affiliation{RTX Technology Research Center (RTRC), East Hartford, CT 06118, USA }
\author{Samuel Berweger}
\affiliation{ National Institute of Standards and Technology, Boulder, CO 80305, USA}
\author{Noah~Schlossberger}
\affiliation{ National Institute of Standards and Technology, Boulder, CO 80305, USA}
\author{Alexandra Artusio-Glimpse}
\affiliation{ National Institute of Standards and Technology, Boulder, CO 80305, USA}
\author{Christopher L. Holloway}
\affiliation{ National Institute of Standards and Technology, Boulder, CO 80305, USA}
\author{Matthew T. Simons}
\affiliation{ National Institute of Standards and Technology, Boulder, CO 80305, USA}

\date{\today}

\begin{abstract}

Rydberg atoms in a gas form are highly sensitive electric field probes capable of detecting and measuring the amplitude, phase, and polarization of broadband time-varying signals.  Here, we present the performance of a frequency modulated continuous wave (FMCW) radar using a Rydberg atom-based subwavelength sensor as a receiver. This sensor down converts the radar echoes, eliminates  key FMCW electrical components, and performs two-dimensional target localization. To demonstrate its capabilities, we present an RF image of a scene containing targets in an anechoic room with radar cross sections down to 0 dBsm at a distance up to 5 m and with a range resolution of 4.7 cm.   

\end{abstract}

\maketitle
Rydberg atom E-field sensors (in hot vapor) are emerging as an alternative technology to the classical receiving antenna for a diverse range of applications \cite{schlossberger2024rydberg, fancher2021rydberg}. These sensors operate by optically exciting atoms to a high principal quantum number ($n \gtrsim 20$) to become
highly sensitive to electric fields, due to the $n^7$ scaling of the polarizability. An optical readout can then determine the frequency, phase, amplitude, and polarization of the radiofrequency (RF) waveform \cite{sedlacek2012microwave, holloway2014, fan2014subwavelength, holloway2017atom, simons2019rydberg, jing2020atomic}, enabling various applications including electric field imaging ~\cite{fan2014subwavelength, schlossberger2024two}, angle-of-arrival detection~\cite{robinson2021determining}, a broadband spectrum analyzer~\cite{meyer2021waveguide}, and temperature measurements~\cite{norrgard2021quantum, schlossberger2025primary}. Even high-bandwidth tasks like real-time video have been demonstrated~\cite{prajapati2022tv}. 

Rydberg atoms have various advantages to their classical counterparts. Classical antennas are fundamentally limited by the Chu Limit \cite{chu1948physical}, which requires the antenna to have several different sizes depending on the band being utilized. In contrast, the Rydberg sensor responds directly to the incident field rather than the power collected, thereby removing any geometric constraints. Rydberg sensors are therefore capable of omnidirectional, ultra-wideband \cite{meyer2021waveguide, zhang2024ultra}, and subwavelength \cite{holloway2014, fan2014subwavelength, holloway2017atom, jau2020vapor} RF sensing enclosed entirely in an all-dielectric form factor. Additionally, while a well matched narrowband antenna will have better sensitivity than a Rydberg atom receiver, the atoms offer an advantage when tunability comes into question. For an unmatched antenna, theoretical analyses indicate the sensitivity of Rydberg atoms will win out~\cite{cox2018quantum, meyer2020assessment}. Finally, the Rydberg sensors may be advantageous for certain receiver applications since they replace electronic components such as RF mixers, antennas (with their size dependent on wavelength), low pass filters, and low noise amplifiers with optical elements such as lasers, fibers, and photodiodes. All of which have the potential to reduce complexity, noise, footprint and weight. 

These benefits of Rydberg sensors provide a unique opportunity for radar applications. A few recent studies have explored Rydberg sensors for continuous wave radar \cite{jing2020atomic}, pulsed radar \cite{bohaichuk2022origins}, satellite radio reflectrometry of soil moisture \cite{arumugam2024remote}, and calibration of automobile millimeter wave radar circuit boards \cite{borowka2024rydberg}. However, FMCW radar has not yet been explored for Rydberg sensors. 

\begin{figure*}[htbp!]
    \includegraphics[width=\linewidth]{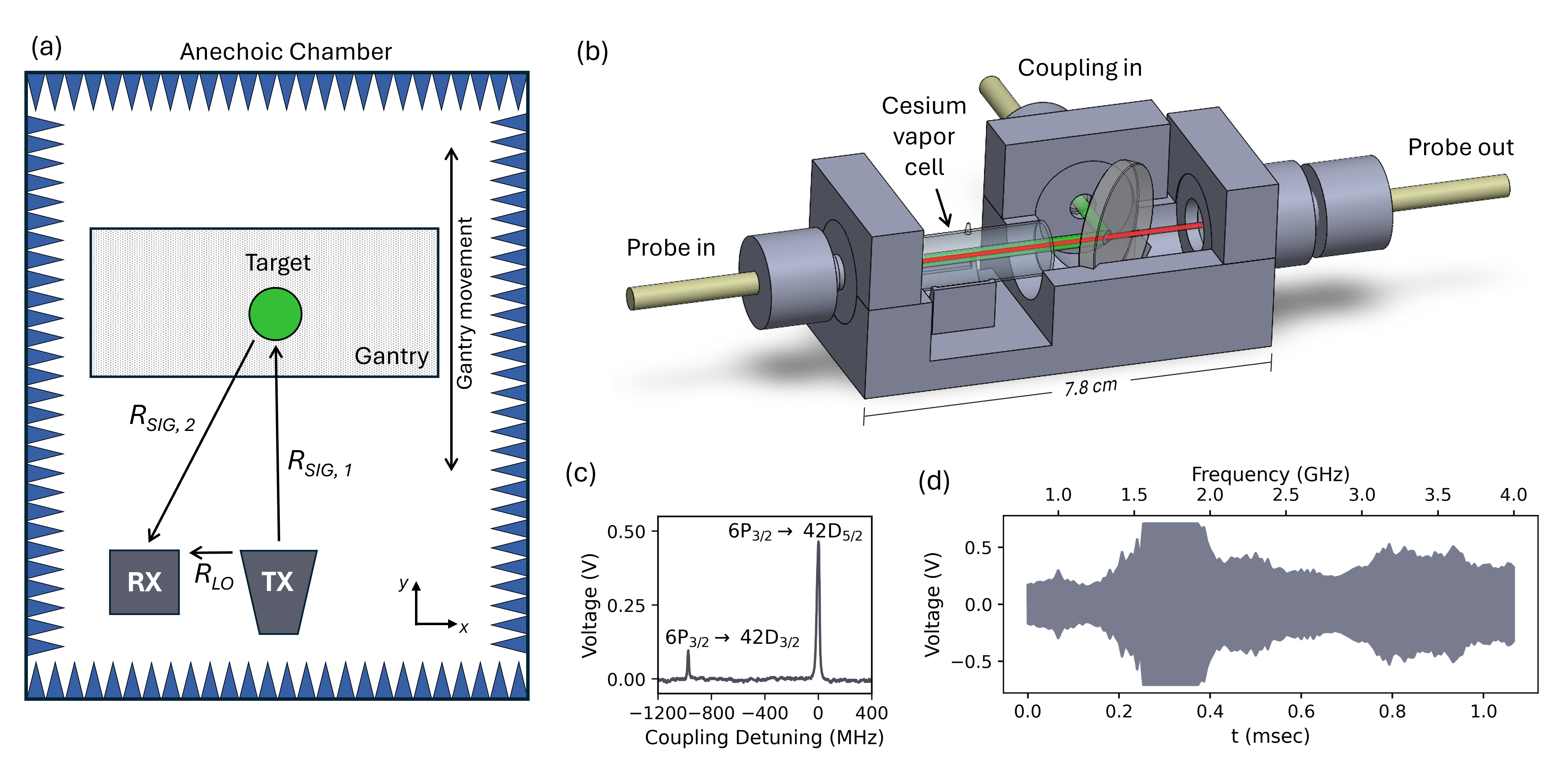}
    \caption{\textbf{Rydberg FMCW radar experimental setup.} (a) An example layout (image not drawn to scale) of one-dimensional target detection in the anechoic chamber. The boresight axis is parallel to the y-axis with the E-field polarized into and out of the page. (b) Our custom designed all-dielectric fiber coupled Rydberg receiver. (c) Electromagnetic induced transparency spectra for locked probe and scanning coupling laser. (d) The transmitted amplitude from the AWG over one FMCW chirp of 1.066 ms  spanning 800~MHz - 4000~MHz.}
    \label{Fig:ExperimentSetup}
\end{figure*}

\begin{figure*}[t]
    \includegraphics[width=0.95\textwidth]{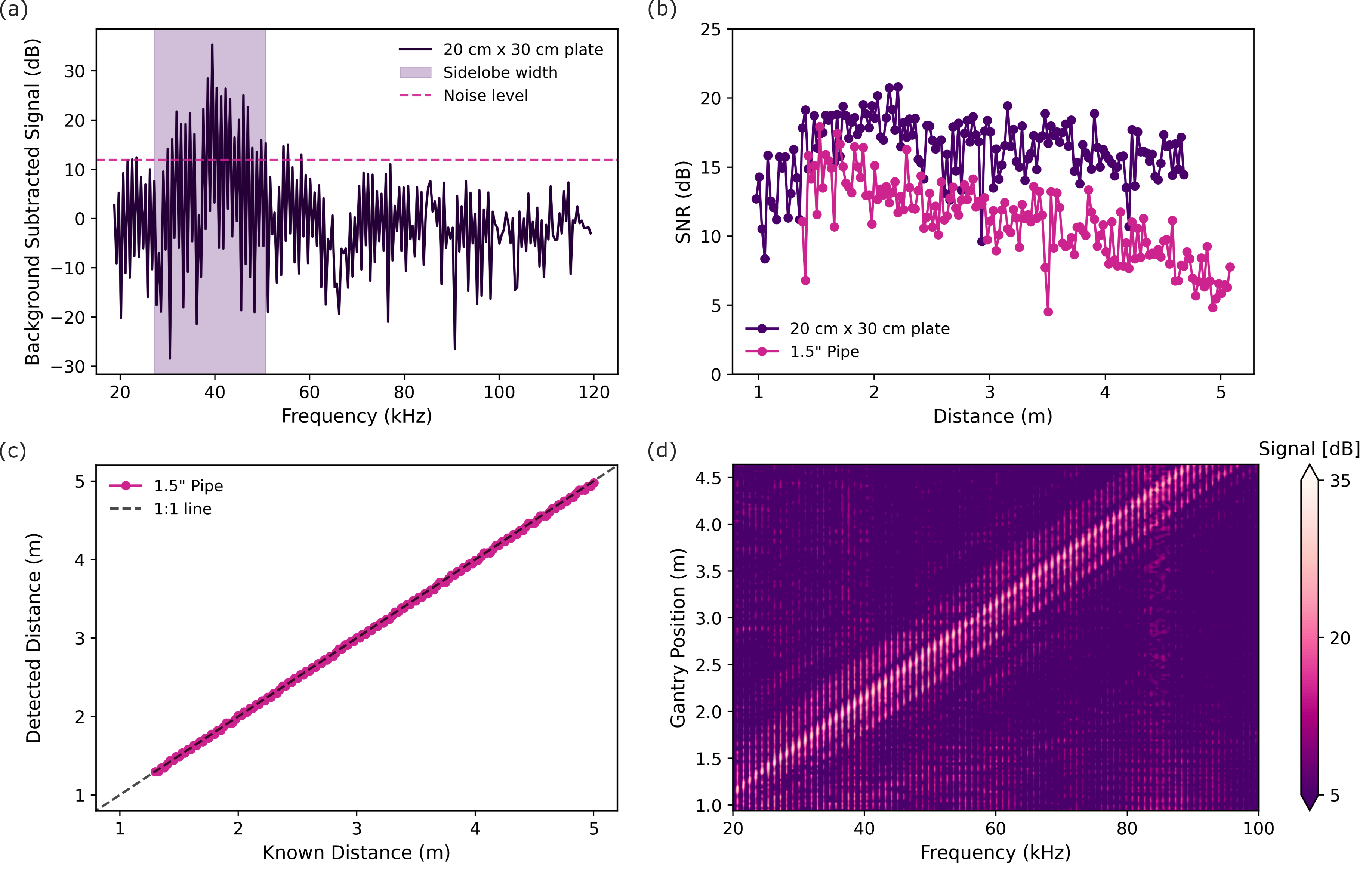}
    \caption{\textbf{Target localization from the Rydberg radar for objects oriented along the transmitting antenna's boresight axis.} (a) A frequency domain signal for a 20 cm x 30 cm plate with the target beatnote near 40 kHz. (b) SNR of a plate and pipe swept longitudinally in the anechoic chamber. The SNR for the plate decreases at distances less than 2 m due to multipath reflections. (c) The location of the 1.5" pipe was correctly identified for all boresight locations tested within the anechoic chamber. (d) A heatmap showing the radar signal of the plate as the gantry position is moved. }
    \label{Fig:RadarData}
\end{figure*}
In this manuscript, we demonstrate an FMCW radar scheme that takes advantage of the benefits of the Rydberg atom sensor. The FMCW radar scheme consists of a single transmitter and a Rydberg receiver that directly mixes the transmitted and reflected signals to generate beat frequencies and thereby determine the target range (Fig. \ref{Fig:ExperimentSetup}a). In this sense, the Rydberg atoms act as a quantum analog to the traditional RF mixer, in which two high frequency signals are down-converted to obtain the intermediate, or beat note, frequency. The RF radiation emitted from the transmit antenna towards the Rydberg receiver serves as the local oscillator (LO) while the radiation reflected from the target creates the signal (SIG). The observed beat note, $f_{beat}$, generated from mixing the LO and reflected SIG for a linear FMCW chirp is given by 
 \begin{equation}
     f_{beat} = \frac{f_{span}}{T_{span}}\frac{R}{c}~,
 \end{equation}
where the distance, $R = R_{SIG,1} + R_{SIG,2} - R_{LO}$ and each component of $R$ is depicted in Fig. \ref{Fig:ExperimentSetup}a, $c$ is the speed of light, $f_{span}$ is the frequency span of the chirp, and $T_{span}$ is the time duration of the frequency chirp. By measuring the beat note frequency, the target distance, $R_{SIG,1}$ can be determined.

We tested this radar in an anechoic chamber measuring 8.5 m x 7.3 m x 4.9 m lined with 91.4 cm pyramidal absorber with $>$35 dB of attenuation from 500 MHz - 40 GHz. The chamber includes an automated gantry on a 6 m rail and a 7-axis robot capable of precise 3-dimensional positioning of targets within the chamber. We placed various targets on the gantry, while the transmitting antenna and Rydberg receiver positions were kept fixed. To acquire the radar images, the gantry position was scanned so that the targets moved relative to the fixed antenna and Rydberg receiver. The gantry was moved in discrete steps so that at each step the probe response was recorded. We used a 750~MHz - 18~GHz broadband dual-ridged horn as the transmitting antenna. As shown in Fig. \ref{Fig:ExperimentSetup}a-b, the Rydberg receiver (Rx) was positioned 30~cm from the edge of the transmitting antenna (Tx) in the H-plane $90^{\circ}$ from boresight and the sensor was oriented so that the laser beam path was parallel to the E-plane. We positioned the Rydberg receiver in the sidelobe in order to reduce the LO power and better detect weak signals.

\begin{figure*}[htbp!]
    \includegraphics[width=0.9\textwidth]{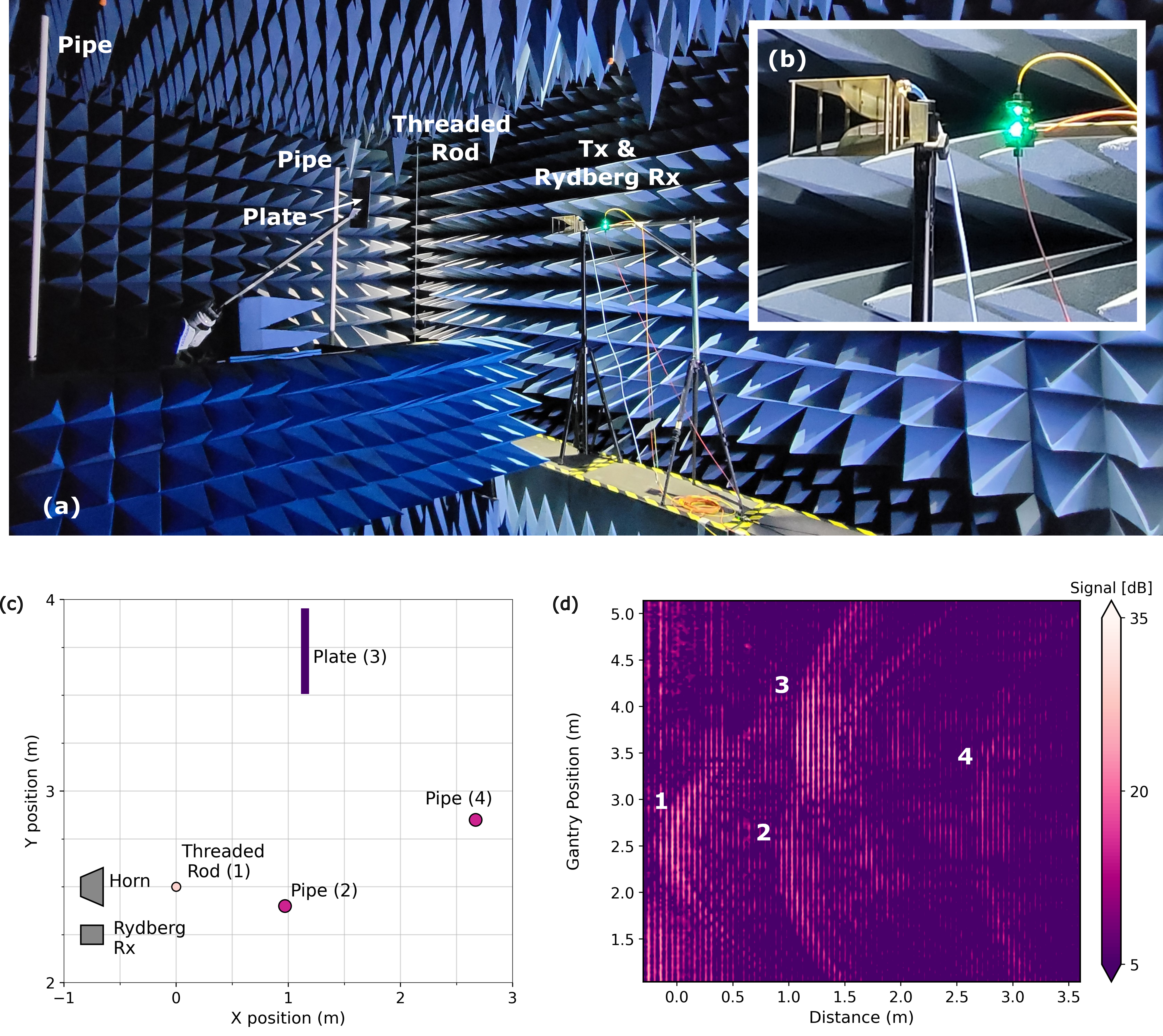}
    \caption{\textbf{Two-dimensional imaging Rydberg radar.} (a) Experimental setup of 4 scattering objects positioned on the movable gantry while the Tx antenna and Rydberg Rx are kept in a fixed location. (b) Close up image of the Tx and Rydberg Rx mounted in the Tx sidelobe. (c)  The spatial coordinates of the objects when the threaded rod is on boresight with the Tx. The 4 scattering objects relative positions with each other do not change throughout the gantry scan. (d) The two-dimensional radar image of the 4 scattering objects. Each object is indicated by a numeric label matched to schematic shown in (c).}
    \label{Fig:Imaging2D}
\end{figure*}

We utilized off-resonant heterodyne detection of cesium atoms excited to a Rydberg state for wideband detection of FMCW chirps. This detection method requires first creating an electromagnetically induced transparency (EIT) in the atomic population through a two photon excitation to a Rydberg state. EIT is described in detail elsewhere \cite{schlossberger2024rydberg, fancher2021rydberg, sedlacek2012microwave, holloway2014}, but briefly, is a process whereby the absorption lines of a probe optical field resonant with a transition from the ground state, $\ket{1}$, to an intermediate state, $\ket{2}$, and a coupling field resonant with a transition $\ket{2}$ to the Rydberg state, $\ket{3}$, become transparent due to a destructive quantum interference in the transition amplitudes from $\ket{1}$ to $\ket{2}$ and $\ket{3}$. Therefore, as the coupling laser is scanned over the $\ket{2} \rightarrow \ket{3}$ transition, the probe transmission changes giving a narrow spectral signature of the Rydberg state. Next, for an applied off-resonant RF field, the energy of the Rydberg state will shift according to the Stark effect by $-\frac{1}{2}\alpha E^2$ where $\alpha$ is the polarizability and $E$ is the field strength. Finally, by locking the probe and coupling frequencies, the down converted beat note between the LO and SIG fields (i.e., heterodyne \cite{simons2019rydberg, jing2020atomic}) can be observed. The experimental specifics are described below.

We performed a two-photon excitation scheme with an 852~nm probe laser locked to the 6S\textsubscript{1/2} (F=-4) $\rightarrow$ 6P\textsubscript{3/2} (F=5) transition and the coupling tuned to 510.101~nm to excite to 42D\textsubscript{5/2}. An example signal while scanning the frequency of the coupling laser is shown in Fig. \ref{Fig:ExperimentSetup}c. We locked the coupling laser off a reference cell external to the anechoic chamber near the maximum field sensitivity location at the half maximum of the positive detuning side of the EIT peak. The atomic response was recorded in the time domain from a single-channel photodetector with 40 dB gain and 90 kHz bandwidth. 

We designed and machined an all-dielectric Rydberg receiver from the thermoplastic polyoxymethylene (Fig. \ref{Fig:ExperimentSetup}b). The main body of the receiver is 7.8~cm in length and holds a cesium vapor cell 10 mm in diameter and 30 mm in length, a longpass dichroic mirror, and three custom fiber collimators. In order to couple the light from our optics lab to the fiber-coupled receiver in the anechoic chamber, we used 30 m of single mode (probe) or 200 $\mu$m core multi-mode (coupling) fibers with a fiber-to-fiber coupler followed by a short 2 m fiber attached directly to the receiver. The input power of the probe into the 2 m fiber (i.e, at the end of the 30 m connecting fiber) was 200 $\mu$W and the input power of the coupling was 200 mW. The input collimators for the probe and coupling lasers include a 5~mm diameter plano-convex lens with 5~mm focal length. The beam diameter for the probe was 1.1~mm while the coupling beam size was 2.2~mm. The probe output featured a 400~$\mu$m core multi-mode fiber and a 10~mm diameter plano-convex lens with 15~mm focal length which went directly to a photodetector. Due to the relatively short path lengths within the receiver, we noticed no difference in the EIT spectrum between a multi-mode versus single-mode fiber and chose the multi-mode fiber because we can easily couple in more optical power. The collection efficiency of the probe laser, measured from the output of the 30 m single mode fiber in the anechoic chamber, to the photodector input, and measured by tuning the laser outside of the cesium D\textsubscript{2} absorption line, was 29\%.

The FMCW chirp signal for radar measurements was generated using an 12~GSa/s arbitrary waveform generator. The chirp was a linear frequency sweep from 800 MHz to 4 GHz with a duration of 1066 $\mu s$ unless otherwise indicated.  We observed variations in RF power at the atoms resulting from the antenna gain factor as a function of frequency. Therefore, we equalized FMCW chirp amplitude in order to correct the changing  LO field strength at the vapor cell location. In general, broadband horns feature a non-linear gain profile. This causes a varying field at the Rydberg receiver for constant transmit input power over the chirp duration. This varying LO field strength as a function of frequency causes the EIT peak to Stark shift independent of any target reflectors -- leading to a major source of transmit/receive leakage noise. To compensate for this, we  equalized the field at the Rydberg receiver using a calibrated dipole field probe \cite{simons2024field}. A transmitted amplitude example after equalization is shown in Fig. \ref{Fig:ExperimentSetup}d.

In order to perform the radar target localization, we measured time-domain signals read directly off the photodetector and performed a fast Fourier transform to measure the beat frequencies. We further performed a background subtraction where the background is measured by removing all targets from the anechoic chamber. An example signal of a 20 cm x 30 cm plate located 2.15 m from the transmitting antenna on boresight is shown in Figure \ref{Fig:RadarData}a. In order to determine the SNR, we assigned a sidelobe width for each reflecting object and took the noise level as the 97.5~\% maximum of the measured trace outside of the sidelobe region. 

The Rydberg FMCW radar system was capable of detecting a 20 cm x 30 cm copper plate and a 3.8 cm x 1 m steel pipe throughout the anechoic chamber (Fig. \ref{Fig:RadarData}b-c). We positioned either the pipe or plate on the boresight axis of the transmitting antenna and automatically moved the gantry as shown in Fig. \ref{Fig:ExperimentSetup}a in 2.5 cm steps. The pipe was oriented with its longitudinal axis parallel to the E-field. The theoretical radar cross section of the plate is $\sigma_{plate}$ = 5 dBsm while the cylinder is $\sigma_{pipe}$ = 0 dBsm. In reality, numerical simulations would be required to get an accurate estimate of $\sigma$ for the near field scattering experiments conducted here. For distances over 3 m, the signal-to-noise ratio (SNR) of the plate versus pipe varies roughly by the difference in the theoretical radar cross sections (Fig. \ref{Fig:RadarData}b). The SNR of the plate for distances less than 2 m decreases because the noise level from multipath reflections increases, as can be seen in the faint line in Fig. \ref{Fig:RadarData}d. Localization of the pipe throughout the room is depicted in Fig. \ref{Fig:RadarData}c. For the simple processing applied to the radar signal here, where the target location is identified by the peak beatnote frequency, the target can be located within a distance, $\Delta R$, set by the FMCW radar range resolution, $\Delta R = c/(2f_{span})$ = 4.7 cm. 

After the initial characterization of the Rydberg FMCW radar's SNR and resolving capabilities, we next demonstrated its applicability to two-dimensional (2D) radar imaging (Fig. \ref{Fig:Imaging2D}a-b). For the 2D imaging experiments, we moved the Tx and Rydberg Rx to the side of the anechoic chamber so that the gantry movement was transverse to the antenna boresight direction (as opposed to the longitudinal scanning illustrated in Fig. \ref{Fig:ExperimentSetup}a). In traditional linear scanning radar, the Tx and Rx (for monostatic radar) are moved with respect to fixed targets. However, due to the constraints of the chamber dimensions we performed a functionally equivalent scenario where the scatterers are moved with respect to fixed antenna positions. We positioned 4 objects on the gantry: two 3.8 cm pipes, a 9.5 mm threaded steel rod, and a 20 cm x 30 cm plate. In this experiment, we used a chirp time duration of 2133 $\mu$s. Similar 2D maps were obtained for 3 scatterers at 1066 $\mu$s chirp duration, but here we show the results for 4 scatterers at 2133 $\mu$s. The relative orientations of the objects are illustrated in Fig. \ref{Fig:Imaging2D}c for a gantry position where the threaded rod is on the boresight axis of the Tx antenna. The 4 scattering objects are clearly resolvable from the 2D scanning image along with the classic hyperbolic scattering profiles typically seen in scanning radar imaging (Fig. \ref{Fig:Imaging2D}d). 

Although radar measurements are standard in classical systems, an atomic Rydberg receiver may provide additional improvements. Here, we demonstrated continuous sensing from 800 MHz to 4 GHz using a single atomic transition in cesium through an off-resonant detection. While similar measurements have been done in waveguide coupled systems ~\cite{meyer2021waveguide}, this is the first demonstration over the air. Specifically related to radar systems, our off-resonant detection scheme has enabled a broadband FMCW radar measurement whereas previous radar measurements have utilized resonant Autler-Townes measurements. For instance, Jing et al. performed resonant heterodyne configuration of CW waveforms capable of measuring non-stationary targets \cite{jing2020atomic} while Bohaichuk et al. \cite{bohaichuk2022origins} described a resonant Autler-Townes field detection to measure the time domain atomic response to a pulsed radar enabling target resolution with 9 m uncertainty. Two additional preprints published during the preparation of this manuscript also proposed resonant Autler-Townes detection of stepped frequency radar \cite{chen2025radar} and a theoretical proposal for FMCW radar signals \cite{cui2025realizing}. In particular, detecting a frequency chirp with resonant detection will inherently be limited by the atomic bandwidth (typically $\sim$10 MHz) with a correspondingly low spatial resolution. Compared to these resonant methods, our broadband FMCW radar scheme enables a higher resolution detection of both stationary and non-stationary targets. 

Some of the main limitations of our Rydberg FMCW radar, as well as many classical systems, concern the resolution and SNR. To improve both the range resolution $\Delta R = c/(2f_{span})$ and the lateral resolution $\Delta l = \sqrt{R c/2f_{span}}$, the frequency span (i.e., bandwidth) must be increased. The transmit bandwidth was limited by our waveform generator with a Nyquist limit of 6 GHz and the amplifier with a band spanning 700 MHz - 6 GHz. The atomic receiver bandwidth is limited to values with relatively small changes in atomic polarizability, i.e., to a frequency less than the RF transition frequency of nearby energy states. For example, for the tested state here, 42D\textsubscript{5/2}, the nearest transition is to 42P\textsubscript{3/2} at an RF frequency of 9.92 GHz.  Thus, going down in principal quantum number can increase the available bandwidth at the expense of reducing the field sensitivity. 

Improving SNR through hardware and signal processing remains an active area of investigation for classical radar. Many of the same issues also affect the Rydberg system, such as thermal noise, clutter, and transmitter-to-receiver leakage. Here, although we did not undergo an extensive examination of noise suppression, we did utilize background subtraction and waveform predistortion to minimize some sources of noise. Interestingly, because the Rydberg system is ideally photon shot noise limited rather than thermal noise limited, future work could achieve lower noise limits than available in classical systems.

In conclusion, we described a Rydberg FMCW radar system and illustrated its use for near-field radar imaging. Our system leveraged a custom designed fiber coupled Rydberg sensor as the receiver in a bistatic radar
configuration. We transmitted linear FMCW chirps and
measured the beat note between the transmitted and reflected signals to perform 2D target localization. In future work, we aim to study the interplay of sensitivity and bandwidth versus principal quantum number, the effect of atomic bandwidth versus SNR, and noise reduction strategies. The work here demonstrates Rydberg FMCW radars could prove consequential for several radar applications including ground penetrating radar, aviation, maritime navigation, and weather forecasting.

\section*{Acknowledgments}
The information, data, or work presented herein was funded in part by the Advanced Research Projects Agency-Energy (ARPA-E), U.S. Department of Energy, under Award Number DE-AR0001859. The views and opinions of authors expressed herein do not necessarily state or reflect those the and of the  United States Government or any agency thereof.
A contribution of the U.S. government, this work is not subject to copyright in the United States.

\subsection*{Conflict of Interest}
\vspace{-3mm}
The authors have no conflicts to disclose.
\vspace{3mm}
\subsection*{Data Availability Statement}
\vspace{-3mm}
All data presented in this work is available to the reader at 

\bibliography{cites}
\end{document}